\documentclass[12pt]{article}
\usepackage{amsmath,amssymb,graphics}

\newlength{\abstractwidth}
\newcommand{\ed}{\varepsilon}
\newcommand{\ap}{\alpha^{\prime}}

\newcommand{\gtab}{\Tilde{G}_{a b}}
\newcommand{\gt}{\Tilde{G}}
\newcommand{\ft}{\Tilde{\mathfrak F}}

\newcommand{\fL}{\mathcal{L}}

\newcommand{\rd}{\dot{r}}

\newcommand{\rds}{\dot{r}^2}

\newcommand{\td}{\dot{t}}

\newcommand{\tds}{\dot{t}^2}
\begin{document}
\renewcommand{\theequation}{\thesection.\arabic{equation}}
\begin{titlepage}
\bigskip
\rightline{} 
 \rightline{hep-th/0107218}
\bigskip\bigskip\bigskip\bigskip
\centerline{\Large \bf {D-branes and Cosmological Singularities}}
\bigskip\bigskip
\bigskip\bigskip

\centerline{\large Brook Williams}
\bigskip\bigskip
\centerline{\em Department of Physics, UCSB, Santa Barbara, CA. 93106}

\begin{abstract}
The motion of a test Dq-brane in a Dp-brane background is studied.
The induced metric on the test brane is then interpreted as the
cosmology of the test brane universe.  One is then able to resolve
the resulting cosmological singularities.  In particular, 
for a D3-brane in a D5-brane background, one finds a 3+1
dimensional FRW universe with equation of state $p = \ed$.  It has
been argued that this may have been the dominant form of matter at
very early times. 

\noindent
\end{abstract}
\end{titlepage}
\baselineskip=18pt \setcounter{footnote}{0}

\section{Introduction}
It has been shown that singularities are generic features of
cosmological and collapse solutions in general relativity.  These
singularities are representative of the breakdown of the classical
theory at short distances.  It is hoped that string theory will allow
us to resolve these singularities, and indeed for many classical
solutions involving time-like singularities this is the case.
In particular, consider the time-like singularities associated with
D-branes.  The geometry outside a D-brane is given by the supergravity (SUGRA)
solution for an extremal black $p$-brane with Q units of R-R charge
(see section 2).
These solutions ($p \ne 3$) are singular.  However, as curvatures
become large the supergravity solutions breakdown and 
stringy effects become important.  One should then replace 
the space-time description with
the description of the gauge theory living on the brane.  Another
interesting example is provided by M-theory.  One
can show that the singular geometry produced by a D6-brane can be
obtained from the dimensional reduction of the a smooth 11
dimensional geometry, analogous to the Kaluza-Klein monopole in 5 dimensions.

There has been much less progress in understanding the space-like
singularities which are present in cosmological models.  A recent idea
which has received a lot of attention is that of an Ekpyrotic Universe 
\cite{ekpy}.  In this model our universe is created by the
collision of a slowly moving bulk brane with the visible brane on which we
(will) live.   Prior to the collision our universe is a cold, empty
place.  Upon collision the branes fuse together creating a ``hot,
thermal bath of radiation and matter'' living on the brane.  The
scenario takes advantage of the non-locality of branes to solve the
horizon problem, supersymmetry to address flatness, and quantum
fluctuations to explain large scale density fluctuations.

In this note a toy model is proposed which also uses coincident branes
to describe the origins of our universe.  A test Dq-brane is allowed to fall
into, or be produced from, a stack of Dp-branes.  The induced metric
on the test brane is then interpreted as the cosmology on the test brane
universe.  Moreover, the cosmology has a space-like singularity at the
point when the branes become coincident.  This differs from the
Ekpyrotic universe in many respects.  Most importantly, in this model the
universe exists before the collision with (or after the production of)
the probe brane.  In Ekpyrotic scenario the ``hot'' universe 
exists only after the bulk brane fuses with the visible brane.  The authors
of \cite{ekpy} used the properties of branes to give new solutions to
many cosmological questions.  In this note our interest is simply to use the
current understanding of how to resolve the time-like singularities associated
with D-branes to learn how string theory might resolve the space-like 
singularities associated with our early universe.  Indeed,  
for a Dp-brane background the singularity induced on the
Dq-brane can be resolved.  Namely, for $\mbox{p} \ne 3$ the proper
picture at short distances, where the induced metric is becoming singular,
is that of the gauge theory living on the brane.  For $\mbox{p}=3$ the
space-time is non-singular and the resulting cosmological singularity
induced on the brane can be understood entirely from the SUGRA perspective.

Section 2 gives a brief review of D-brane metrics and some details
about how the cosmology of the test brane will be interpreted.
Section 3 will discuss a Dp-brane moving in a Dq-brane
background and section 4 applies the results of section 3 to the
specific case of a D3-brane moving in D5-brane background.

\section{D-brane Metrics}
The geometry produced by flat D-branes is homogeneous and isotropic parallel
to the brane and spherically symmetric in the directions transverse to
the brane.  Therefore the induced metric on a test brane, which lies
parallel to the D-brane, should also be homogeneous and isotropic, and
is thus described by a FRW cosmology.                          
The metric for Q Dp-branes sitting at the origin is given by \cite{jbbs},
$$
ds^2 = Z(r)^{-1/2} \eta_{\mu \nu} dX^\mu dX^\nu +  Z(r)^{1/2} dX^mdX^m \ \ .
$$
\begin{eqnarray*}
\mbox{Here} \ \ \ e^{2 \phi}   &=&   Z(r)^{(3-p)/2} \\ 
                 Z(r)   &=&   1 +\frac{\rho^{7-p}}{r^{7-p}} \\
     r^2 \equiv X^m X^m & & 
          \rho^{7-p}  \propto \alpha^{\prime (7-p)/2} \ g \ Q \ \ .
\end{eqnarray*}
The indices  $\mu, \nu$ correspond to directions along the brane
while $m, n$ run transverse to the brane.  Throughout this paper
we will use the gauge, 
$\xi^0 = \tau \ $  $\ \xi^i = X^i$.
Here $\xi^a$ are the coordinates on the brane.  Since the branes are
parallel the indices $a,b$ are a subset of the $\mu, \nu$.  $\tau$ will
be chosen so that the
induced metric is in the standard FRW form, i.e. $\gt_{0 0} = -1$.
The motion of the test brane is assumed to be radial (in the transverse
space) and lie along $X^{p+1}, \ (r \equiv |X^{p+1}|)$.  Furthermore,
it will be assumed that $Q \gg 1$ so that the back reaction of the test
brane on the geometry can be ignored.  
The induced metric on the Dq-brane, $\gtab = G_{M N} 
\frac{\partial X^M}{\partial \xi^a}
\frac{\partial X^N}{\partial \xi^b}$, is therefore,
$$
\gtab = Z^{-1/2}
\begin{pmatrix}
-\tds + \rd^2 Z &                                              \\
                 & \mbox{\Huge $\mathbf{1}$\normalsize$_{(q \times q)}$} 
\end{pmatrix} \ .
$$
We will find that it is possible to make the gauge choice 
$\gt_{0 0} = -1$, which can be written as,
\begin{equation}
Z^{-1/4} \sqrt{\tds - \rds Z} = 1 \ .
\label{goo}
\end{equation}
This brings the induced metric to that of a flat Robertson Walker geometry,
$$ ds^2 = - d\tau^2 + a^2(\tau)dX^idX^i \ \ .$$
Here $a(\tau) = Z^{-1/4}[r(\tau)]$.
$\gtab$, will be interpreted as the cosmology on
the $q+1$ dimensional universe.
  
At this point we shall assume the naive perspective of a $q+1$
dimensional classical observer, and thus interpret the expansion of
the brane universe as if it were due to a perfect fluid content in standard
Einstein gravity.  It is worth noting that in the toy models discussed below
there is no real matter living on the brane.  Thus the expansion of
the universe,
which is actually being caused by the curvature of the higher
dimensional space, is incorrectly perceived to be the result of dark matter
living on the brane. 

\section{Solving Equations of motion}
The motion of a Dq-brane moving in a Dp-brane background is governed by
the DBI action\cite{jbbs},
\begin{equation}
S_q = -\mu_q \int d^{q+1}\xi \ e^{- \phi} \sqrt{-det(\gtab + \ft_{a b})}
\ + \mu_q \int_{q+1} \Tilde{C}_{q+1} \ .
\label{DBI}
\end{equation}
For simplicity it will be assumed that there is no gauge field living
on the test brane and that there is no antisymmetric closed string
background living in the bulk; thus $\ft_{a b} = 0$.  
For parallel branes if $p \ne q$ then $\Tilde{C}_{q+1}$, the pullback
of the Ramond-Ramond potential, vanishes.  However, for 
$p=q$ \cite{johnsn}\footnote{In \cite{johnsn} $\Tilde{C}_{q+1}$ is
given in static gauge.  Here we have changed coordinates 
$\xi^0 = X^0 \rightarrow \xi^0 = \tau$.},
$$ 
\Tilde{C}_{q+1} = (Z_p^{-1} - 1) \ \td \
d\tau \wedge dx^1 \wedge \ldots \wedge dx^q \ \ .
$$
Our action is then,
\begin{equation}
S_q = -\mu_q \int d^{q+1}\xi \ \Bigl[Z^{(p-q-4)/4} 
\ \bigl(\sqrt{\tds - Z \rds} \ - \ \delta_{p,q} \ \td \ \bigr) 
+ \delta_{p,q} \ \td \Bigr] \ .
\label{action}
\end{equation}

We can now find the classical equations of motion.  Defining
$s \equiv -(p-q-4)/4$ Lagrangian takes the form, 
$$
\fL  = Z^{-s} ( \sqrt{ \tds - Z \rds } - \td \delta_{s,1})+\td \delta_{s,1} \ .
$$
Since $t(\tau)$ is a cyclic variable we know that 
$\partial \fL / \partial \td = c + \delta_{s,1}$, 
where $c$ is a constant of the motion\footnote{Here we have added
$\delta_{s,1}$ to $c$ in order to simplify (\ref{tdot}).}.  Taking
the derivative and solving for $\tds$ gives,
\begin{eqnarray}
\tds = \frac{(Z^{s}c + \delta_{s,1})^2 Z \rds}{l_{p-q}^2} \ ,
\label{tdot}
\end{eqnarray}
where we have defined $l_{p-q}^2 \equiv (Z^{s} c + \delta_{s,1})^2 - 1$.
The gauge choice (\ref{goo}) relates $r(\tau) \ \mbox{and} \ t(\tau)$.
Therefore the equation of motion coming from the variation of $\fL$
with respect to
$r(\tau)$ doesn't provide any new information.  From (\ref{goo})
and (\ref{tdot}) it is easy to see that,
\begin{equation}
d \tau = \pm \frac{Z^{1/4}}{l_{p-q}} dr \
\label{dtau}
\end{equation}
One can now apply these results to various cases.  The specific
example of a D3-brane moving in a D5-brane background yields an
interesting cosmology and is discussed below.

\section{An Example:  D5-brane Background}

Let's now consider the motion of a D3-brane in a D5-brane background.   
For this case (\ref{dtau}) can be integrated exactly.  The integral
over $r$, however, gives rise to quartic roots and hypergeometric
functions which cannot be algebraically inverted to give $r(\tau)$.
Fortunately the interesting behavior, namely the big bang/big crunch
takes place when the branes are nearly
coincident.  In this limit, $r \ll \rho$, we can see that,
\begin{eqnarray*}
d \tau \simeq \pm \frac{Z^{-1/4}}{c} dr \simeq
\frac{1}{c}\sqrt{\frac{r}{\rho}} dr\\
\Longrightarrow \frac{r(\tau)}{\rho} = \mathcal{C}^2
(\frac{\tau}{\rho})^{2/3}, \ \ \mbox{where}
\ \ \mathcal{C}^2 \equiv \frac{2}{3 c} \ .
\end{eqnarray*}
Here we have assumed $c \ge 0$, which is required for the induced metric
to be Lorentzian.
The induced metric on the brane is then,
$$
ds^2 = - d\tau^2 + \mathcal{C}^2 (\frac{\tau}{\rho})^{2/3} dX^i dX^i \ .
$$
A scale factor $a(\tau) \propto \tau^{1/3}$ corresponds to a perfect
fluid with the equation
of state $p = \ed$.  It has been argued by Banks and
Fischler\cite{bankfish} that this may have been the dominant form of
matter in the early stages of our universe.  The argument is the
following:  $p = \ed$ is the
stiffest equation of state such that the velocity of sound is less
than or equal to the speed of light.  
Furthermore, Fischler and Susskind argued that for a FRW cosmology this
is the stiffest equation of state satisfying the holographic 
bound\cite{fs}.  These are taken as indications that 
this is the stiffest equation of state allowed by
nature.  For a FRW universe, with perfect fluid and 
$p = \gamma \ed$, 
Einstein's equations give $\ed = a( \tau )^{-3(1+\gamma)}$.  So if
there exists some $p = \ed$ matter, and this is the stiffest
equation of state allowed by nature, then this form of matter will dominate 
at early times.

There might be concern that the limit $r \ll \rho$ is
outside the region of validity of the supergravity solutions we are
using. A careful treatment of this was done in \cite{imsy}.  It was
found that the SUGRA solutions could be trusted in the region
\begin{equation}
\frac{1}{\sqrt{g Q}}  \ll \frac{r(\tau)}{\sqrt{\ap}} \ll
\sqrt{\frac{Q}{g}} \ .
\label{imsy}
\end{equation}
Note that for large $(g Q)$, which is required since the
back-reaction from the test brane on the geometry has been ignored,
the SUGRA solutions hold very close to the brane.
Similarly, as the induced geometry on the brane becomes singular there
are corrections to the DBI action which become important.  Thus the
above solution can only be trusted as long as
the induced scalar curvature on the brane, $R$, is much less than
the string scale,  
$R \ll 1 / \alpha^{\prime}$.  For our test brane universe 
$R \propto \tau^{-2}$.
This can be seen by direct computation or by dimensional
arguments.  From this we can see $\alpha^{\prime} \ll \tau^2$, 
which provides a lower bound on $r$.  It follows that,
\begin{equation}
\mathcal{C}^2 (g Q)^{1/6} 
\ll \frac{r(\tau)}{\sqrt{\ap}} \ll \sqrt{g Q} \ .
\label{53valid}
\end{equation}
The upper bound comes from working in the region, $r \ll \rho$.
Note that the lower bound can be taken to zero by choosing
$\mathcal{C} \sim 0$.  This limit corresponds to having the test brane
initially moving along an almost null path.
Our solution is valid in the region which both (\ref{imsy}) and (\ref{53valid})
are satisfied.

The breakdown of the SUGRA solution marks the end of our classical
knowledge.  String theory, as previously mentioned, now tells us that 
the correct physical picture
is that of the gauge theory living on the brane.  Any system for which
the number of $ND$ coordinates\footnote{the number of coordinates such that a
string stretched between the branes has Neumann boundary conditions on
one end and Dirichlet boundary conditions on the other} is equal to 2 has
an open string tachyon which is not projected out in the GSO
projection.  As the
D3-brane approaches the D5-branes these tachyons will form and thus
prevent the D3-brane from passing through the D5-branes and escaping
to infinity.  Rather, the tachyon indicates an instability whose
end result is the D3-brane dissolving into the D5-branes and then spreading
out as magnetic flux on the
branes.  For a detailed discussion see \cite{cnds}.  In
addition to the magnetic flux there
will be ripples produced from the momentum of the D3-brane.
Observers living on the D3-brane falling into the D5-branes,
will thus find their universe collapsing, and then suddenly, the 3+1
dimensional universe to which they are accustomed becomes a 5+1
dimensional universe.  Sadly, this shocking development does not save our
poor observers.  Rather they find themselves being spread out as
magnetic flux and
ripples in their new universe.  Alternatively, and perhaps
more importantly, a D5-brane
with incoming flux and ripples could, under very fine tuned initial
conditions, cause the production of a D3-brane, giving birth to
our universe.
\section{Discussion}

Understanding the origins of our universe should be a central part of the
``theory of everything''.  String theory, 
if it is such a theory, should certainly be able to resolve the cosmological
singularities associated with birth of our universe. 
In the above note we have discussed a toy model in which cosmological
singularities can be resolved in string theory.  The case of a
D3-brane moving in a
D5-brane background was discussed in detail.  For such a scenario one
is able to give a complete description of the apparent singularity.  As the
geometry on the probe-brane becomes
singular the space-time description of the brane universe must be
reinterpreted as the gauge theory
living on the D5-branes;  our 3+1 dimensional universe dissolves into the
D5-brane and spreads out as magnetic flux and ripples on the brane.  
The equation of state of this universe was found to be $p = \ed$, 
which, as discussed in
section 4, may have been relevant at very early stages of our universe.  

These are interesting results, but is this a realistic model?
Certainly not.  For starters we have not included any matter on the
brane universe, which one might argue is unrealistic.  Including
matter would complicate the analysis but in
principle it could be done.  More importantly we have not confined
gravity to the brane.  Though the model discussed does not give a
realistic description of our
universe it does give us some insight into how string theory might
resolve the space-like singularities associated with our early universe.

There are other resolutions of singularities which may be obtained by looking at
different background geometries.
In particular, consider that of a D3-brane moving in a D3-brane background.
One finds that this background does not yield a physical equation
of state.  However, the resolution of the singularity is 
very interesting;   
the cosmological singularity can be understood as arising from the motion
of the probe brane through a perfectly smooth space-time!  
Moreover, the singularity on a D4-brane\footnote{By  dimensional
reduction one is still able to obtain a model for a 3+1 dimensional universe.}
moving in the presence of a D6-brane could be understood by lifting up to 11
dimensions, where the geometry is non-singular, thus providing an
M-theoretic resolution to cosmological
singularities.  Recall, however, that in order to avoid complications
caused by back-reaction on the metric we have taken the $Q$, the number
of D6-branes, to be very large.  For multiple D6-branes the 11
dimensional geometry contains a
conical singularity.  One could consider a similar set up with a
single D6-brane.  In such a scenario, however, complications
would arise from the back reaction of the metric by the D4-brane,
and perhaps more importantly, the breakdown of SUGRA
approximations. 

\vskip 1.25cm \centerline{\bf Acknowledgments} \vskip 1cm

I would like to thank Gary Horowitz for discussion and guidance
throughout this project.  I would also like to thank Simeon Hellerman and
Veronika Hubeny for giving careful readings and useful feedback of an
early draft, and Alex Flournoy for discussion.
Work supported in part by NSF grant PHY00-70895.

\vskip 1.25cm \centerline{\bf \large Note Added \normalsize} 
\vskip 1cm

After completing this work it was pointed out that much of what has
been discussed here is contained within a paper, ``Mirage Cosmology'',
by A. Kehagias and E. Kiritsis \cite{mirage}.

\pagebreak

\end{document}